\documentclass[12pt]{iopart}
\usepackage{epsfig,graphics}
\begin{document}

\title{Chiral $SU(3)$ Symmetry and Strangeness}

\author{M F M Lutz\dag,
E E  Kolomeitsev\ddag, C L  Korpa\P }

\address{\dag\ Gesellschaft f\"ur Schwerionenforschung (GSI),\\
Planck Str. 1, D-64291 Darmstadt, Germany\\ Institut f\"ur
Kernphysik, TU Darmstadt\\ D-64289 Darmstadt, Germany}
\address{\ddag\ ECT$^*$, Villa Tambosi, I-38050 \,Villazzano  (Trento) \\
and INFN, G.C.\ Trento, Italy}
\address{\P\ Department of Theoretical Physics, University of
Pecs, \\Ifjusag u.\ 6, 7624 Pecs, Hungary}

\begin{abstract}

In this talk we review recent progress on the systematic evaluation of
the kaon and antikaon spectral functions in dense nuclear matter based
on a chiral SU(3) description of the low-energy pion-, kaon- and antikaon-nucleon
scattering data.

\end{abstract}



\ead{Matthias F.M. Lutz: m.lutz@gsi.de}

\section{Introduction}

A good understanding of the antikaon spectral function in nuclear matter is required for the
description of $K^-$-atoms \cite{Gal,Florkowski} and the subthreshold production of kaons in heavy ion
reactions \cite{Senger}. An exciting consequence of a significantly reduced effective
$K^-$ mass could be that kaons condense in the interior of neutron stars \cite{K:condensation:1,K:condensation:2,Brown-Lee}.
The ultimate goal is to relate the in-medium spectral function of
kaons with the anticipated chiral symmetry restoration at high baryon density.
To unravel quantitative constraints on the kaon spectral functions from subthreshold kaon production data
of heavy-ion reactions requires transport model calculations which are performed
extensively by various groups \cite{Brown-Lee,Fuchs,Aichelin,Bratkovskaya,Ko}. The next generation of
transport codes which are able to incorporate  more consistently particles with finite width
are being developed \cite{Knoll,Leupold,Cassing,Schaffner}. This is of considerable
importance when dealing with antikaons which are poorly described by a quasi-particle
ansatz \cite{ml-sp,ramossp}.

There has been much theoretical effort to access the properties of
kaons in nuclear matter
\cite{Lutz,Pethick,Kolomeitsev,Waas1,Waas2,ml-sp,ramossp,Tolos}.
An antikaon spectral function with support at energies smaller
than the free-space kaon mass was already anticipated in the 70's
by the many K-matrix analyses of the antikaon-nucleon scattering
process (see e.g. \cite{Martin}) which predicted considerable
attraction in the subthreshold scattering amplitudes. This leads
in conjunction with the low-density theorem \cite{DoverHuf,Lutz}
to an attractive antikaon spectral function in nuclear matter.
Nevertheless, the quantitative evaluation of the antikaon spectral
function is still an interesting problem. The challenge is first
to establish a solid understanding of the vacuum antikaon-nucleon
scattering process, in particular reliable subthreshold
antikaon-nucleon scattering amplitudes are required, and secondly,
to evaluate the antikaon spectral function in a systematic
many-body approach.

The antikaon-nucleon scattering is complicated due to the open inelastic
$\pi \Sigma$ and $\pi \Lambda $ channels and the presence of the s-wave $\Lambda(1405)$
and p-wave $\Sigma(1385)$ resonances just below and the d-wave $\Lambda(1520)$ resonance
not too far above the antikaon-nucleon threshold. In this talk we review recent
progress obtained within the newly formulated $\chi$-BS(3) approach,
for chiral Bethe-Salpeter approach to the SU(3) flavor group \cite{LuKol}.
It constitutes a systematic and non-perturbative application of the chiral SU(3) Lagrangian
to the meson-baryon scattering problem consistent with covariance, crossing
symmetry, large-$N_c$ sum rules of QCD and the chiral counting concept. The low-energy
pion-, kaon- and antikaon-nucleon scattering data were reproduced successfully demonstrating
that the chiral SU(3)flavor symmetry is a powerful tool to analyze and predict
hadron interactions systematically. The amplitudes obtained in that scheme are particularly
well suited for an application to the nuclear kaon dynamics, because it was demonstrated
that they are approximately crossing symmetric in the sense that the $K N$ and $\bar K N$
amplitudes smoothly match at subthreshold energies. Therefore we believe that those
amplitudes, which are of central importance for the nuclear kaon dynamics, lead to
reliable results for the propagation properties of kaons in dense nuclear
matter \cite{LuKor}.

As was pointed out in \cite{ml-sp} the realistic evaluation of the antikaon self energy
in nuclear matter requires a self consistent scheme. In particular the feedback effect of
an in-medium modified antikaon spectral function on the antikaon-nucleon scattering process was
found to be important for the $\Lambda(1405)$ resonance structure in nuclear matter.
In this talk we present a selection of results obtained in a novel covariant many-body
framework \cite{LuKor}. Self consistency was implemented in terms of the free-space meson-nucleon
scattering amplitudes, where the amplitudes of the $\chi$-BS(3) approach were used.
Besides presenting realistic kaon and antikaon spectral functions we discuss the in-medium
structure of the s-wave $\Lambda (1405)$ and p-wave $\Sigma (1385)$ resonances.

\section{Kaon- and antikaon-nucleon scattering}

We briefly review the most striking phenomena arising when applying
the chiral $SU(3)$ Lagrangian to the kaon- and antikaon-nucleon interaction
processes. At leading chiral order these interactions are supposedly described
by the Weinberg-Tomozawa term,
\begin{eqnarray}
{\mathcal L}_{WT } &=&
\frac{i}{8\,f^2}
( \bar N \,\gamma_\mu \, N) \big( (\partial^\mu\,K)^\dagger \,K
-K^\dagger (\partial^\mu \,K )\big)
\nonumber \\
&+&\frac{3\,i}{8\,f^2}
( \bar N \,\gamma_\mu\,\vec{\tau} \, N)
\big( (\partial^\mu\,K )^\dagger \,\vec{\tau}\,K
-K^\dagger \vec{\tau} \,( \partial^\mu\,K ) \big) \,,
\label{WT}
\end{eqnarray}
where the parameter $f \simeq f_\pi \simeq 92.4$ MeV is known from the
decay process of charged pions. In contrast to the successful application
of the chiral Lagrangian in the flavor $SU(2)$ sector, its application
to the strange sector of QCD is flawed by a number of subtleties if the
rigorous machinery of chiral perturbation theory is applied. The leading
interaction term (\ref{WT}) fails miserably in describing the s-wave
scattering lengths of both kaons and antikaons off a nucleon. Most
stunning is the failure of reproducing the repulsive $K^- p$
scattering length \cite{Iwasaki}. The chiral Lagrangian
predicts an attractive scattering length at leading order instead.
This is closely linked to the presence of the $\Lambda(1405)$ resonance
in the $K^-p$ scattering amplitude just below the $K^-p$ threshold. Considerable
theoretical progress has been made over
the last few years by incorporating the dynamics of that $\Lambda(1405)$ resonance
into the chiral dynamics. The key point is to change approximation strategy and
expand the interaction kernel rather than the scattering amplitude directly. That
amounts to solving some type of coupled-channel scattering equation like the
Lippmann-Schwinger or the Bethe-Salpeter equation. As a consequence the
$\Lambda (1405)$ resonance  is generated dynamically by coupled channel effects.
A realistic description of the antikaon-nucleon scattering process requires
the inclusion of all SU(3) channels $\bar K N, \pi \Lambda, \pi \Sigma , \eta \Sigma,
\eta \Lambda$ and $K \Xi$ together with correction terms predicted by
the chiral $SU(3)$ Lagrangian. The number of free parameters controlling the chiral
correction terms can be significantly reduced by insisting on sum rule relations as
they arise from QCD in the large-$N_c$ limit. For a detailed description of an
up-to-date scheme with a more complete list of references we refer to the recent
work \cite{LuKol}. In that work it is shown that the chiral $SU(3)$ Lagrangian does
describe all low-energy
pion-, kaon- and antikaon-nucleon scattering data fairly well, once chiral perturbation
theory is applied to the covariant scattering kernel of the Bethe-Salpeter scattering
equation.

In this talk we would like to discuss an aspect in more detail, which is particularly
important when applying the chiral kaon- and antikaon-nucleon dynamics to kaon and
antikaon propagation in dense nuclear matter. Crossing symmetry relates the $\bar K N$
and $K N$ scattering amplitudes at subthreshold energies. This manifests itself in
the form of the dispersion-integral representation of the antikaon-nucleon scattering
amplitude, ${\rm T}^{(0)}_{\bar K N}(\sqrt{s}\,)$, evaluated in forward direction,

\begin{figure}[t]
\begin{center}
\includegraphics[width=8cm,clip=true]{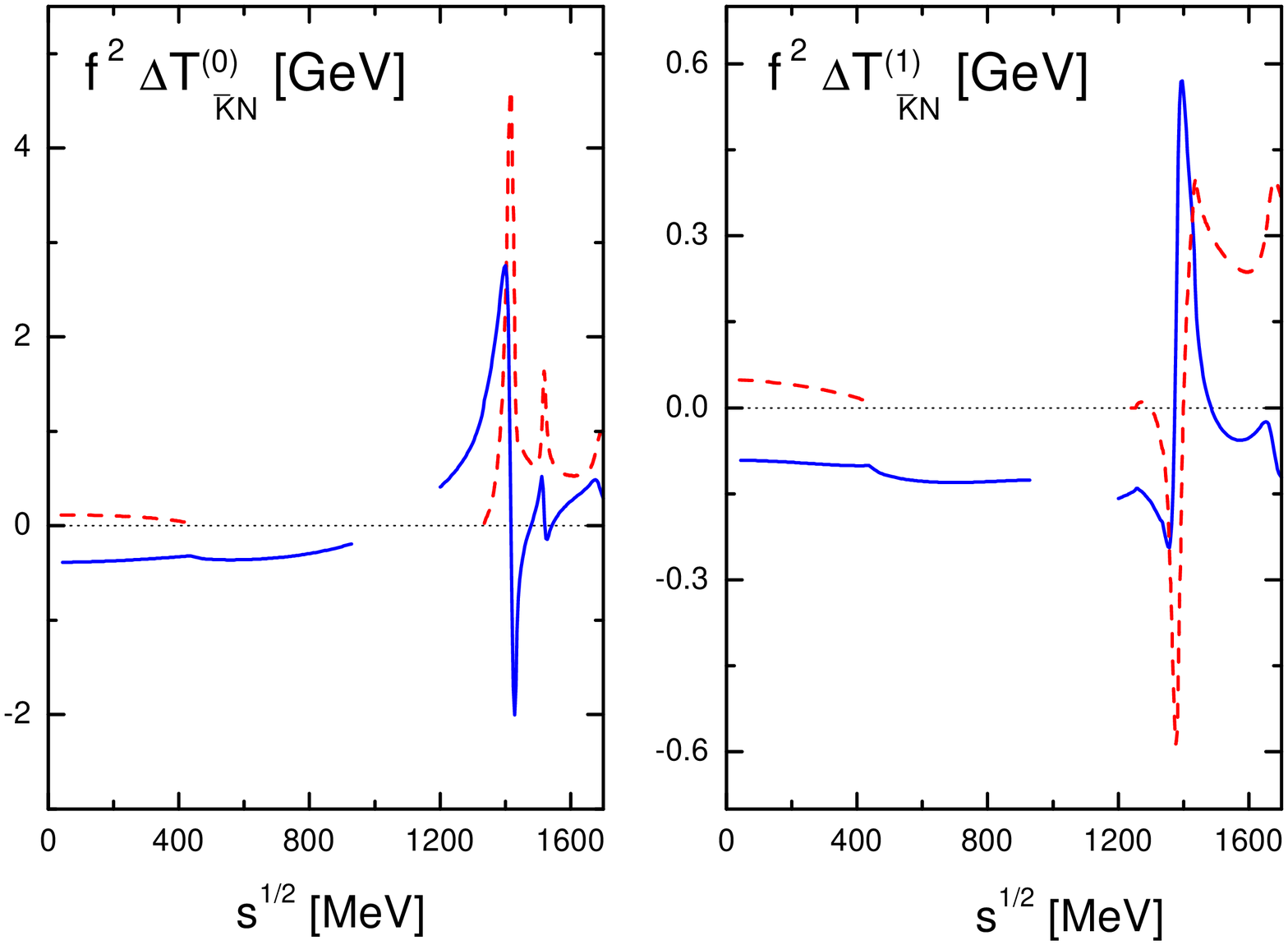}
\end{center}
\caption{Approximate crossing symmetry of the hyperon-pole subtracted kaon-nucleon forward
scattering amplitudes. The lines in the left hand parts of the figures result from the
$KN$ amplitudes. The lines in the right hand side of
the figures give the $\bar K N$ amplitudes.}
\label{fig:crossing}
\end{figure}

\begin{eqnarray}
{\rm T}^{(0)}_{\bar K N}(\sqrt{s})-{\rm T}^{(0)}_{\bar K N}(\sqrt{s_0}) &=&
\frac{f^2_{KN \Lambda }}{s-m^2_\Lambda}-\frac{f^2_{KN \Lambda }}{s_0-m^2_\Lambda}
\nonumber\\
&+&\int_{-\infty}^{(m_N-m_K)^2} \frac{d \,s'}{\pi }\,\frac{s-s_0}{s'-s_0}\,
\frac{\Im \,{\rm T}^{(0)}_{\bar K N} (\sqrt{s'})}{s'-s -i\,\epsilon}
\nonumber\\
&+&\int_{(m_\Sigma+m_\pi)^2}^{+\infty} \frac{d \,s'}{\pi }\,\frac{s-s_0}{s'-s_0}\,
\frac{\Im \,{\rm T}^{(0)}_{\bar K N} (\sqrt{s'})}{s'-s -i\,\epsilon} \;,
\label{cross-disp-check}
\end{eqnarray}
where we performed one subtraction at $s=s_0$ to help the convergence of the dispersion integral.
For sake of clarity we consider here the isospin zero channel only.
The scattering amplitude ${\rm T}^{(0)}_{\bar K N}(\sqrt{s})$ shows unitarity cuts
not only for $\sqrt{s}>m_\Sigma+m_\pi$, representing for instance the inelastic process
$\bar K N \to \pi \Sigma $, but also for $\sqrt{s}<m_N-m_K$ reflecting
the elastic $K N$ scattering process.

A problem arises because the kaon- and
antikaon-nucleon scattering processes are described by two distinct Bethe-Salpeter equations.
If the Bethe-Salpeter interaction kernel of the $\bar K N$ sector, that implies a
particular scattering amplitude ${\rm T}^{(0)}_{\bar K N}(\sqrt{s})$, is evaluated in
perturbation theory, the amplitude ${\rm T}^{(0)}_{\bar K N}(\sqrt{s})$ does not describe
properly the unitarity cuts of the $KN$ channel, manifest at the subthreshold energy
$\sqrt{s}< m_N-m_K$. Analogously, an evaluation of the kaon-nucleon scattering amplitudes
in terms of a perturbative interaction kernel is not reliable far
below threshold where the $\bar K N$ channel opens. Of course, crossing symmetry
is reconciled once the interaction kernels of the $K N$ and $\bar K N$ sectors include
the appropriate infinite class of Feynman diagrams. It is important to face this problem since
the in-medium antikaon spectral function tests the subthreshold $\bar K N$ amplitudes,
\begin{eqnarray}
\Pi_{\bar K} (\omega , \vec q=0) \simeq  \Big( {\textstyle{1\over 4}}\,T^{(0)}_{\bar K N}(\omega +m_N) +
{\textstyle{3\over 4}}\,T^{(1)}_{\bar K N}(\omega +m_N)\Big) \, \rho +\cdots
\end{eqnarray}
where we recalled the low-density theorem written down for for the antikaon polarization function
$\Pi_{\bar K} (\omega ,\vec q)$ \cite{DoverHuf,Lutz}. If an antikaon can propagate
in dense nuclear matter with $\omega < m_K$ and say $\vec q=0$ for simplicity, the evaluation
of the antikaon polarization function $\Pi_{\bar K}(\omega ,0)$ at that energy requires the
knowledge of the scattering amplitudes $T^{(I)}_{\bar K N}(\sqrt{s})$ at a subthreshold
energy $\sqrt{s}< m_N+m_K$.

This problem is solved in the chiral coupled channel framework by supplementing the
Bethe-Salpeter equation with an appropriate renormalization program that leads to the
matching of the subthreshold $K N$ and $\bar K N$ amplitudes,
\begin{eqnarray}
&&  T^{(0)}_{\bar K N}(s) \simeq
-\frac{1}{2}\,  T^{(0)}_{K N}(2\,s_0-s)+\frac{3}{2}\, T^{(1)}_{ K N}(2\,s_0-s) \;,
\nonumber\\
&&   T^{(1)}_{\bar K N}(s) \simeq
+ \frac{1}{2}\,  T^{(0)}_{K N}(2\,s_0-s)+\frac{1}{2}\,  T^{(1)}_{ K N}(2\,s_0-s) \;,
\label{exp-cross}
\end{eqnarray}
close to the optimal matching point $s_0=m_N^2+m_K^2$. This is a crucial input of
the $\chi$-BS(3) approach developed in \cite{LuKol}. Even though the renormalization
of the chiral Lagrangian in perturbation theory is straightforward, this it not anymore
the case once the infinite diagram summation implied by the Bethe-Salpeter equation is
performed. An additional renormalization condition is needed. In \cite{LuKol} it was strongly
argued that this condition is naturally provided by the crossing symmetry constraint.
In Fig.~\ref{fig:crossing} we demonstrate that the hyperon pole-subtracted scattering amplitudes,
$ \Delta T^{(I)}_{K N}$ and $  \Delta T^{(I)}_{\bar KN} $, comply with the crossing
identities (\ref{exp-cross}) close to $s \simeq s_0=m_N^2+m_K^2$ approximatively. We are
therefore convinced that our subthreshold antikaon-nucleon scattering amplitudes are
determined reliably and well suited for an application to the nuclear kaon dynamics.

\begin{figure}[t]
\begin{center}
\includegraphics[width=8cm,clip=true]{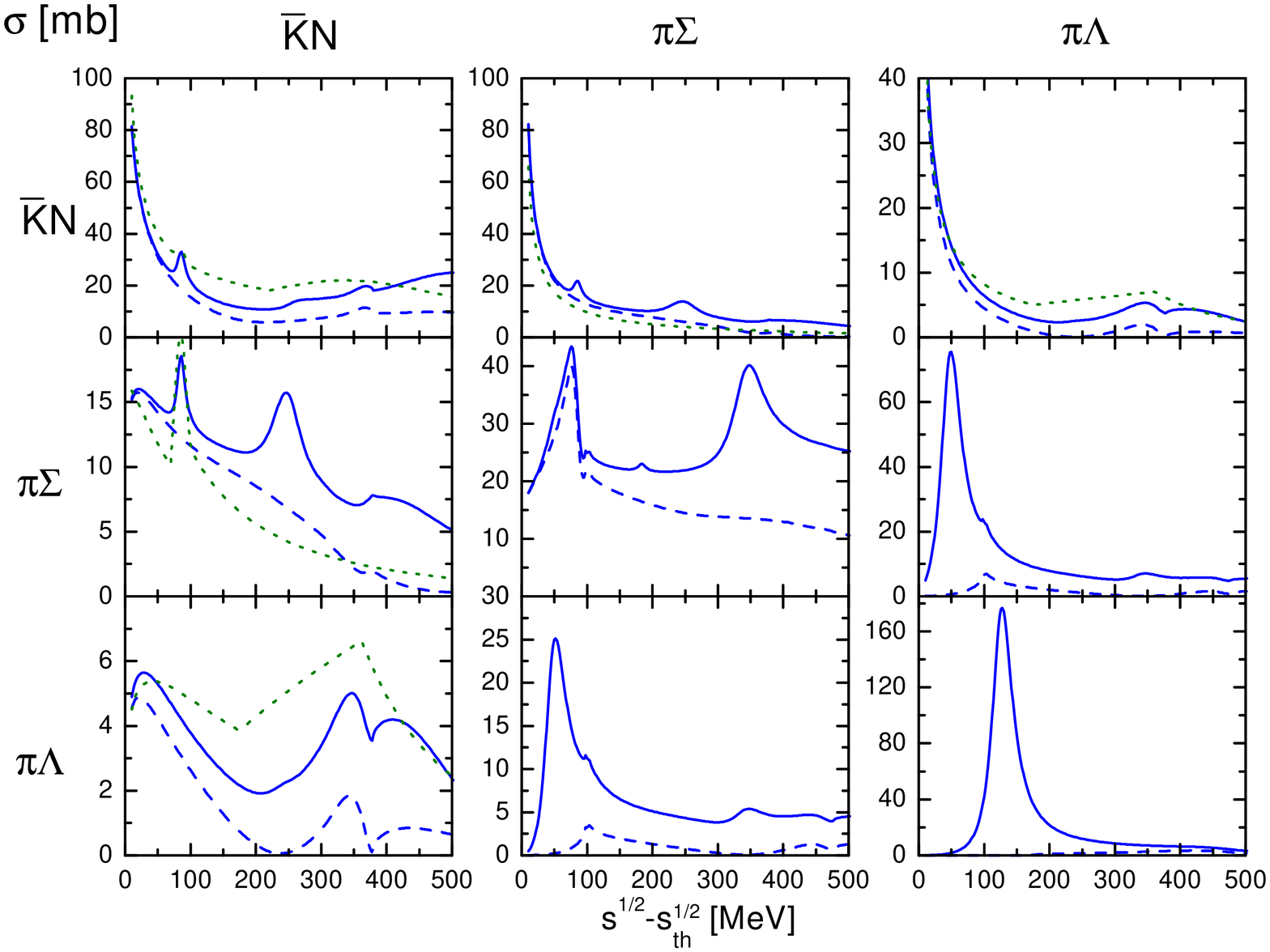}
\end{center}
\caption{Total cross sections
$\bar K N \to \bar K N$, $\bar K N \to \pi \Sigma $, $\bar K N \to \pi \Lambda $ etc
relevant for subthreshold
production of antikaons in heavy-ion reactions. The solid and dashed lines give the
results of the $\chi$-BS(3) approach
with and without p- and d-wave contributions respectively. The
dotted lines correspond to the parameterizations given in \cite{Brown-Lee}.}
\label{fig:cross-pred}
\end{figure}

We close this section on the microscopic input by a presentation of total cross sections
relevant for transport model calculation of heavy-ion reactions. We believe that the
$\chi$-BS(3) approach is particularly well suited
to determine some cross sections not directly accessible in scattering experiments.
Typical examples would be
the $\pi \Sigma \to \pi \Sigma , \pi \Lambda$ reactions. Here the quantitative realization of
the chiral
SU(3) flavor symmetry including its important symmetry breaking effects are an extremely useful
constraint
when deriving cross sections not accessible in the laboratory directly. It is common to consider
isospin averaged cross sections \cite{Cugnon,Brown-Lee}
\begin{eqnarray}
\bar \sigma_{}(\sqrt{s}\,) = \frac{1}{N}\,\sum_{I} \,(2\,I+1)\,\sigma_{I}(\sqrt{s}\,) \,.
\end{eqnarray}
The reaction dependent normalization factor is determined by $ N= \sum \,(2\,I+1)$ where the
sum extends over isospin channels which contribute in a given reaction. In
Fig. \ref{fig:cross-pred} we confront the cross sections of the channels
$\bar K N, \pi \Sigma $ and $\pi \Lambda$ with typical parameterizations used in transport
model calculations. The cross sections in the first column are determined by detailed balance
from those of the first row. Uncertainties are present nevertheless, reflecting the large
empirical uncertainties of the antikaon-nucleon cross sections close to threshold. The
remaining four cross sections in Fig. \ref{fig:cross-pred} are true predictions of the
$\chi$-BS(3) approach. We should emphasize that we trust our results quantitatively only for
$\sqrt{s}< 1600$ MeV. It is remarkable that nevertheless our cross sections agree with the
parameterizations in \cite{Brown-Lee} qualitatively up to much higher energies except in the
$\bar K N  \leftrightarrow \pi \Sigma $ reactions where we overshoot those parameterizations
somewhat. Besides some significant deviations of our results from \cite{Cugnon,Brown-Lee} at
$\sqrt{s}-\sqrt{s_{\rm th}} <$ 200 MeV, an energy range where we trust our results
quantitatively, we find most interesting the sizeable cross section of about 30 mb for
the $\pi \Sigma \to \pi \Sigma $ reaction. Note that here we include the isospin two
contribution as part of the isospin averaging. As demonstrated by the dotted line in
Fig. \ref{fig:cross-pred}, which represent the $\chi$-BS(3) approach with s-wave contributions
only, the p- and d-wave amplitudes are of considerable importance for the
$\pi \Sigma \to \pi \Sigma $ reaction.

\section{Kaon and antikaon propagation in nuclear matter}

\begin{figure}[t]
\begin{center}
\includegraphics[width=6cm,clip=true]{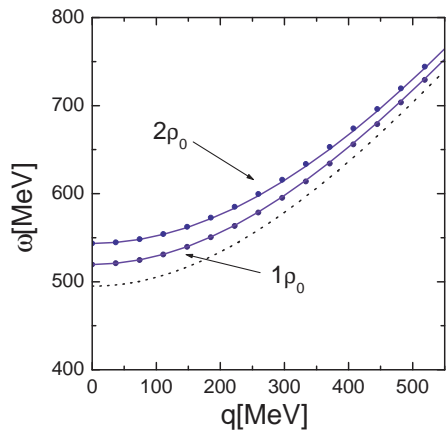}
\end{center}
\caption{The spectrum of a kaon with energy $\omega $ and momentum $\vec q$
at nuclear saturation density, $\rho_0$, and $2\,\rho_0$ . The solid
line follows from the real part of the $KN$ scattering amplitudes
of~\cite{LuKol}. The circles represent the parameterization~(\ref{kplus:spec})
and the dashed line shows the free-space kaon spectrum for comparison.}
\label{fig:kplus}
\end{figure}

We begin with a discussion of kaon propagation in dense nuclear matter.
The kaon self energy is evaluated in terms of the real parts of the
s- and p-wave kaon-nucleon scattering amplitudes of \cite{LuKol}.
The resulting quasi-particle energy $E_K(\vec q\,)$ is shown in
Fig. \ref{fig:kplus} for nuclear saturation density, $\rho_0$, and 2 $\rho_0$.
We neglect small rescattering effects proportional to $(T_{KN})^n$ with $n\geq 2$, which
introduce a 15 $\%$ correction term at small kaon momenta only. A nuclear environment
leads to an increase of the kaon energy where the effect is reduced as the kaon momentum
$\vec q $ increases. Our result can be interpreted rather accurately
in terms of scalar and vector potentials, parameterizing the kaon polarization
function,
\begin{eqnarray}
\Pi_K(\omega , \vec q) \simeq \Big(1.1\,m_K-\omega +0.2\,\frac{\vec q^2}{m_K}\Big) \,
46.8\,{\rm MeV} \,\frac{\rho}{\rho_0} \,,
\label{kplus:spec}
\end{eqnarray}
for $\omega = E_K(\vec q\,)$. We emphasize that the parameterization
of the polarization function $\Pi_K(\omega , \vec q)$ is reliable only at the quasi-particle
energy of the kaon, $E_K(\vec q)$, as is implied by (\ref{kplus:spec}). In particular, the
parameterization (\ref{kplus:spec}) does not describe correctly any derivatives of the
polarization function. The parameterization $\Pi_K(\omega , \vec q)$ is in striking
contradiction to the naive representation,
\begin{eqnarray}
\Pi_K(\omega , 0) = \Big(\frac{3\,\omega }{4f^2}- \frac{\Sigma_{KN}}{f^2}\,\Big) \,\rho \,,
\label{kplus:WT}
\end{eqnarray}
in terms of the Weinberg-Tomozawa term and the so called kaon-nucleon sigma term,
$\Sigma_{KN}> 0$, frequently seen in the literature. The scalar and vector terms in
(\ref{kplus:spec}) and (\ref{kplus:WT}) have opposite signs. As we argued in
the previous section, the chiral Lagrangian does not describe the kaon-nucleon scattering
process correctly if evaluated in perturbation theory. The small attractive energy dependence of
the self energy reflects important range terms required to describe the s-wave KN phase shifts.
The rather small repulsive momentum dependence follows from the net-repulsion of the p-wave
amplitudes. For non-zero momentum the small attractive vector potential in
(\ref{kplus:spec}) leads to a reduction of the repulsion implied by the s-wave
scattering lengths and p-wave scattering volumes. We checked that the parameterization
(\ref{kplus:spec}) is valid to high accuracy for kaon momenta smaller than $|\vec q |< 600$ MeV
reproducing quantitatively the kaon quasi-particle energy $E_K(\vec q)$. This
is clearly demonstrated in Fig.~\ref{fig:kplus} where we compare the spectrum
following from the exact polarization operator with that one from the parameterization
(\ref{kplus:spec}) at two different nuclear densities.

\begin{figure}[t]
\begin{center}
\includegraphics[width=14cm,clip=true]{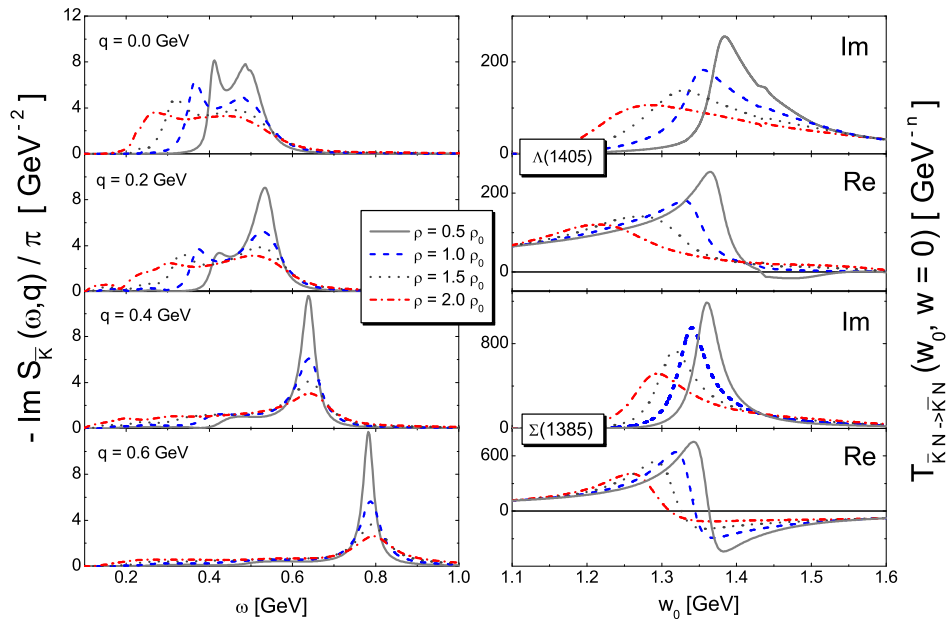}
\end{center}
\caption{The antikaon spectral function is shown in the left hand panel as
a function of the antikaon energy $\omega$, the momentum $q$ and the
nuclear density with $\rho_0 = 0.17$ fm$^{-3}$. The right hand panel
illustrates the in-medium modification of the  $\Lambda(1405)$ and
$\Sigma (1385)$ hyperon resonances. It is plotted the real and imaginary
parts of the antikaon-nucleon scattering amplitudes in the appropriate channels.
The hyperon energy and momentum are $w_0$ and $w =0$ respectively.}
\label{fig:kaon-sp}
\end{figure}

We turn to antikaon and hyperon resonance propagation in dense nuclear matter. In
Fig. \ref{fig:kaon-sp} we present the antikaon spectral function together with
the antikaon-nucleon scattering amplitudes of selected channels at various nuclear
matter densities, $\rho $, as evaluated in a self consistent manner in \cite{LuKor}.
The antikaon spectral function exhibits a rich structure with a pronounced dependence
on the antikaon three-momentum. That reflects the coupling of the $\Lambda (1405)$
and $\Sigma (1385)$ hyperon states to the $\bar K N$ channel. Typically the peaks
seen are quite broad and not always of quasi-particle type. As was emphasized
in \cite{ml-sp,LuKor} the realistic evaluation of the antikaon propagation in nuclear
matter requires the simultaneous consideration of the hyperon resonance propagation.
The most important contributions, the s-wave $\Lambda (1405)$ and p-wave $\Sigma (1385)$
resonances, experience important medium modifications as demonstrated in Fig.
\ref{fig:kaon-sp}.
The results at $2 \,\rho_0$ should be considered cautiously because nuclear binding
and correlation effects were not yet included in \cite{LuKor}.

\section{Summary}

We reviewed the application of the microscopic $\chi$-BS(3) dynamics developed recently in
\cite{LuKol,LuKor} to kaon, antikaon and hyperon resonance propagation in nuclear matter.
Of central importance for the microscopic evaluation of the kaon and antikaon spectral
functions in cold nuclear matter are the kaon- and antikaon-nucleon scattering amplitudes,
in particular at subthreshold energies. The required amplitudes are well established
by the $\chi$-BS(3) approach and show sizeable contributions from p-waves not considered
systematically so far \cite{Kolomeitsev,ml-sp,ramossp}. For the antikaon spectral function
one finds a pronounced dependence on the three-momentum of the antikaon reflecting
the presence of hyperon-nucleon-hole states. The quantitative evaluation of the antikaon
self energy requires the self consistent consideration of the in-medium change of the
hyperon resonance structures. At nuclear saturation density we reported attractive
mass shifts for the $\Lambda(1405)$, $\Sigma (1385)$ and $\Lambda(1520)$ of about 60 MeV,
60 MeV and 100 MeV respectively. The resonance widths increase to about 120 MeV, 70 MeV
and 90 MeV. Whereas the kaon spectral function is well approximated by
a quasi-particle approach the antikaon spectral function shows typically
a rather wide structure invalidating a simple quasi-particle description.

\end{document}